\documentclass[sigconf,screen]{acmart}

\usepackage{adjustbox}
\usepackage{multirow}
\usepackage{enumitem}

\AtBeginDocument{%
  \providecommand\BibTeX{{%
    \normalfont B\kern-0.5em{\scshape i\kern-0.25em b}\kern-0.8em\TeX}}}

\copyrightyear{2023}
\acmYear{2023}
\setcopyright{rightsretained}
\acmConference[SIGIR-AP '23]{Annual International ACM SIGIR Conference on Research and Development in Information Retrieval in the Asia Pacific Region}{November 26--28, 2023}{Beijing, China}
\acmBooktitle{Annual International ACM SIGIR Conference on Research and Development in Information Retrieval in the Asia Pacific Region (SIGIR-AP '23), November 26--28, 2023, Beijing, China}
\acmDOI{10.1145/3624918.3625332}
\acmISBN{979-8-4007-0408-6/23/11}

\begin{document}

\title{A Comparative Study of Training Objectives for Clarification Facet Generation}
\author{Shiyu Ni}
\orcid{0009-0001-7965-7771}
\affiliation{%
  \institution{CAS Key Lab of Network Data\\Science and Technology, ICT, CAS}
  \institution{University of Chinese Academy of Sciences}
  \city{Beijing}
  \country{China}
}
\email{nishiyu23z@ict.ac.cn}

\author{Keping Bi}
\orcid{0000-0001-5123-4999}
\affiliation{%
  \institution{CAS Key Lab of Network Data\\Science and Technology, ICT, CAS}
  \institution{University of Chinese Academy of Sciences}
  \city{Beijing}
  \country{China}}
\email{bikeping@ict.ac.cn}

\author{Jiafeng Guo}
\orcid{0000-0002-9509-8674}
\authornote{Jiafeng Guo is the corresponding author.} 
\affiliation{
	\institution{CAS Key Lab of Network Data\\Science and Technology, ICT, CAS}
	\institution{University of Chinese Academy of Sciences}
	\city{Beijing}
	\country{China}
}
\email{guojiafeng@ict.ac.cn}

\author{Xueqi Cheng}
\orcid{0000-0002-5201-8195}
\affiliation{%
  \institution{CAS Key Lab of Network Data\\Science and Technology, ICT, CAS}
  \institution{University of Chinese Academy of Sciences}
  \city{Beijing}
  \country{China}}
\email{cxq@ict.ac.cn}

\renewcommand{\shortauthors}{Shiyu Ni, et al.}

\begin{abstract}
Due to the ambiguity and vagueness of a user query, it is essential to identify the query facets for the clarification of user intents. Existing work on query facet generation has achieved compelling performance by sequentially predicting the next facet given previously generated facets based on pre-trained language generation models such as BART. Given a query, there are mainly two types of training objectives to guide the facet generation models. One is to generate the default sequence of ground-truth facets, and the other is to enumerate all the permutations of ground-truth facets and use the sequence that has the minimum loss for model updates. The second is permutation-invariant while the first is not. In this paper, we aim to conduct a systematic comparative study of various types of training objectives, with different properties of not only whether it is permutation-invariant but also whether it conducts sequential prediction and whether it can control the count of output facets. To this end, we propose another three training objectives of different aforementioned properties. For comprehensive comparisons, besides the commonly used evaluation that measures the matching with ground-truth facets, we also introduce two diversity metrics to measure the diversity of the generated facets. Based on an open-domain query facet dataset, i.e., MIMICS, we conduct extensive analyses and show the pros and cons of each method, which could shed light on model training for clarification facet generation. The code can be found at \url{https://github.com/ShiyuNee/Facet-Generation}.

\end{abstract}



\maketitle

\section{INTRODUCTION}
Since user queries can be ambiguous or vague, query intent clarification is beneficial to enhance user experience and retrieval effectiveness. In today's dialogue-based retrieval systems, where the display space or voice bandwidth is limited, it becomes even more important. For example, the query ``Chicago'' can mean ``the musical Chicago'', ``the band Chicago'', ``the city Chicago'', and ``the movie Chicago'', etc. Accurately predicting such query facets can be conducive to various search scenarios. In web search, displaying possible facets can help users refine their original queries or provide them with relevant subtopics that they may find useful. In a conversation search system, facets can be used to ask clarifying questions.

Early studies on extracting query facets mainly rely on specific domains or external resources \cite{dakka2008automatic, kohlschutter2006using, latha2010afgf, li2010facetedpedia, stoica2007automating, liu2023topic}. However, such methods may not be suitable for open-domain queries, which have a much more variety of facets. Later, some studies \cite{dou2011finding, kong2013extracting, kong2014extending, kong2016precision} find that utilizing top-retrieved documents can help identify open-domain query facets based on the word frequencies \cite{dou2011finding} and their co-occurrences \cite{kong2013extracting, kong2014extending, kong2016precision}. Nevertheless, these methods could not capture semantic relations sufficiently. 
In recent years, grounded on pre-trained language models (PLMs) such as BERT \cite{devlin2018bert} and BART \cite{lewis2019bart}, query facet prediction has achieved compelling performance \cite{zamani2020generating,hashemi2021learning, hashemi2022stochastic, samarinas2022revisiting}. Although query facet prediction can be formulated as facet generation, sequence labeling, facet classification, etc., facet generation has more flexibility than the others and has been found to perform the best \cite{samarinas2022revisiting}. In this paper, we focus on facet generation based on PLMs with top-retrieved documents.  

Existing methods typically conduct facet generation by generating the sequence of facets associated with a query \cite{hashemi2021learning, samarinas2022revisiting, hashemi2022stochastic}, using two common training objectives where the ordering of a facet sequence has an impact or no impact during training. 
One is to generate the default sequence of facets \cite{hashemi2021learning, samarinas2022revisiting}, denoted as \textit{Seq-Default} in Table \ref{tab:different objectives}. This approach forces the model to generate a specific facet ordering that is only one of the possible combinations, which may lead to suboptimal performance. Realizing this issue, \citet{hashemi2022stochastic} propose the other training objective that enumerates the permutations of the query facets and uses the one with minimum loss to guide model training, denoted as \textit{Seq-Min-Perm} in Table \ref{tab:different objectives}. Theoretically, the training objective should be permutation-invariant since the ground-truth facets are an unordered set. However, enumerating the permutations incurs huge computation costs and only using minimum loss may not leverage the permutations of facets sufficiently. When we examine the existing methods from other perspectives \cite{hashemi2021learning, samarinas2022revisiting,hashemi2022stochastic} (See Table \ref{tab:different objectives}), we find that they all generate the next facet depending on the previously generated facets. Also, the count of generated facets is determined by the model and no more facets can be generated if the model has already output a termination symbol. 

 \begin{table*}[htbp]
\caption{Comparison of characteristics among different objectives. $\checkmark$ means true and $\times$ means false. $|q|$ is the average length for all the queries, $|d|$ is the average length for all the concatenations of documents, $m$ is the average number of facets for each query and $|f|$ is the average length for all the facets. $\mathcal{A}_m^i$ means selecting $i$ elements without repetition from $m$ elements, considering the ordering.}
  \label{tab:different objectives}
  \scalebox{0.76}{
  \begin{tabular}{lcccc}
    \toprule
    Model & Sequential-Prediction & Permutation-Invariant & Facet-Count-Controllable & Complexity-Per-Training-Epoch \\
     \midrule
 Seq-Default & $\checkmark$ & $\times$ & $\times$ & $O(n*((|q| + |d|)^2 + (m*|f|)^2 + (|q| + |d|) * m*|f|))$  \\
  Seq-Min-Perm & $\checkmark$ & $\checkmark$ & $\times$ & $O(n*m!*((|q| + |d|)^2 + (m*|f|)^2 + (|q| + |d|) * m*|f|)) $ \\
 Seq-Avg-Perm & $\checkmark$ & $\checkmark$ & $\times$ & $O(n*m!*((|q| + |d|)^2 + (m*|f|)^2 + (|q| + |d|) * m*|f|))$ \\
 Set-Pred & $\times$ & $\checkmark$ & $\checkmark$ & $O(n*m*((|q| + |d|)^2 + |f|^2 + (|q| + |d|) * |f|))$ \\
 Seq-Set-Pred &  $\checkmark$ & $\checkmark$ & $\checkmark$ & $O (n*(\sum_{i=0}^{m-1}\mathcal{A}_m^i*\mathcal{A}_{m-i}^1)*((|q| + |d| + i*|f|)^2 + |f|^2 + (|q| + |d| + i*|f|) * |f|))$ \\
    \bottomrule
  \end{tabular}
  }
\end{table*}   

In this paper, we aim to conduct a comparative study of various training objectives for the task of clarification facet generation from multiple perspectives. To this end, we propose another three training objectives for facet generation that are all permutation-invariant but have different properties in terms of whether to sequentially predict facets based on the context of so-far generated facets and whether the count of generated facets is controllable. 1) \textit{Seq-Avg-Perm}: It is similar to \textit{Seq-Min-Perm} but uses the average loss of facet permutations instead of minimum; 2) \textit{Set-Pred}: It predicts each facet in the ground-truth set as independent targets during training and uses sampling algorithms such as beam search to select the top k facets with the highest probabilities for inference; 3) \textit{Seq-Set-Pred}: It sequentially predicts the remaining facet set based on arbitrarily generated facets as context. As shown in Table \ref{tab:different objectives}, among these methods, only Set-Pred does not output the next facet based on previously generated facets. So, it does not have the training cost due to enumerating the facet permutations. However, it could suffer from keep generating similar facets due to only referring to the original query and top documents. Only Set-Pred and Seq-Set-Pred can output the designated number of facets. They focus on the facet prediction task alone, which relieves them from also concerning how many facets they need to produce and could lead to potentially better performance.  

Previous work mainly evaluates the effectiveness of facet generation by measuring the matching degree between the generated facets and the ground truth facets \cite{hashemi2021learning, samarinas2022revisiting,hashemi2022stochastic}. This, however, can not reflect the diversity of generated facets sufficiently. To conduct systematic comparisons of different types of methods, we introduce two diversity metrics that measure the term and semantic diversity of the generated facets respectively. 
We evaluate models trained with the two existing objectives and three new objectives using both the commonly used matching metrics and our proposed diversity metrics. 
Based on MIMICS \cite{zamani2020mimics}, an open-domain query facet dataset, Our experimental analyses demonstrate that the appropriate permutation-invariant objectives can help generate better facets; Facet prediction that is only based on the query and top-retrieved documents (i.e., Set-Pred) achieves compelling performance in terms of the metrics measuring matching with the ground truth but have the worst diversity performance; Methods that only learn facet prediction given context (i.e., Seq-Set-Pred) have better semantic matching metrics with ground truth but worse diversity performance than its counterpart that also learns when to stop generating facets (i.e., Seq-Avg-Perm). 
To sum up, the main contributions of this work include:

1) To the best of our knowledge, this is the first work that conducts a systematic comparative study on a wide variety of training objectives for clarification facet generation.

2) We propose three extra training objectives that are of different properties from the existing work and introduce two diversity-oriented metrics for evaluation. 

3) We conduct comprehensive analyses and show the pros and cons of each type of method, which we believe could provide insights for future research efforts in clarification facet generation. 

\section{RELATED WORK}
Clarifying user intent is a crucial issue in information retrieval. There have been many works focusing on learning facet prediction. Additionally, some studies utilize predicted intents to generate clarifying questions, clarifying intent by asking questions to the user. So the two threads of research related to our work are: asking clarifying questions and facet prediction.

\subsection{Asking Clarifying Questions}
Asking clarifying questions(CQ) is an important way to clarify the intent of a query \cite{keyvan2022approach}. Typically, before asking a CQ, we need to determine a facet and generate based on the facet. Current research on CQ can be mainly divided into two categories: (1) ranking/selecting CQ as \cite{rao2018learning, aliannejadi2019asking, aliannejadi2020convai3, bi2021asking} and (2) generating CQ like \cite{rao2019answer, zamani2020generating, dhole2020resolving, wang2021template, sekulic2021towards}. 

For ranking/selecting CQ, Rao and Daum{\'e} III \cite{rao2018learning} sorted questions based on the usefulness of their answers, using the Expected Value of Perfect Information as the theoretical basis for ranking. Later, Aliannejadi et al. \cite{aliannejadi2019asking} collected a dataset for CQ and proposed a baseline for selecting CQ. \cite{hashemi2020guided,bi2021asking} conducted CQ ranking based on top-retrieved documents and negative feedback, respectively. 

For CQ generation, Rao and Daum{\'e} III \cite{rao2019answer} applied generative adversarial learning techniques when training the sequence-to-sequence question generation model. Zamani et al. \cite{zamani2020generating} proposed a rule-based model and two neural question generation models to generate CQ when given a query and its facet. Later, Dhole \cite{dhole2020resolving} proposed a model that utilizes rule-based systems to generate discriminative questions, aiming to obtain clarifications on user intent. Wang and Li \cite{wang2021template} introduced a template-based question generation model called TG-ClariQ, which selects a template question from a candidate set. The missing parts in the template question are filled in using selected words. They converted generating CQ into a selection task. Meanwhile, Sekuli{\'c} et al. \cite{sekulic2021towards} also proposed a facet-driven approach and Zhao et al. \cite{zhao2022generating} showed that such facets can be extracted from top retrieved documents. Recently, Wang et al. \cite{wang2023zero} converted CQ generation to a facet-constrained question generation task to guide effective and precise question generation.  

\subsection{Facet Prediction}
Current research on facet prediction can be mainly divided into two categories: (1) \textbf{facet extraction} and (2) \textbf{facet generation}.

The majority of early work on facet extraction \cite{dakka2008automatic, kohlschutter2006using, latha2010afgf, li2010facetedpedia, stoica2007automating} primarily relied on specific domains or external resources.  Kohlschtter et al. \cite{kohlschutter2006using} introduced an approach for extracting based on personalized PageRank link analysis and annotated taxonomies. Stoica et al. \cite{stoica2007automating} put forward a technique for generating hierarchical faceted metadata. The method utilized hypernym relations in WordNet to extract this metadata from textual descriptions of items.
Subsequently, Dakka and Ipeirotis\cite{dakka2008automatic} use entity hierarchies in Wikipedia and WordNet to extract candidate facet terms. Li et al. \cite{li2010facetedpedia} created a faceted retrieval system that showcases pertinent facets extracted from Wikipedia hyperlinks and categories. While these methods have shown promising results in certain scenarios, they often struggle under large-scale open-domain settings.
Apart from the approaches mentioned above that rely on specific domains or external resources, another approach for facet extraction and generation is based on top retrieved documents. Dou et al. \cite{dou2011finding} introduced QDMiner, one of the earliest open-domain facet extraction systems. This system utilizes textual patterns to aggregate frequent lists from the top web search results and gets query dimensions based on the aggregated results. Kong and Allan \cite{kong2013extracting} proposed a graph-based probabilistic model for determining whether a candidate term is a facet term and for identifying whether two candidate terms belong to the same query facet. Later, they extended faceted search to the general web \cite{kong2014extending}. In their subsequent work \cite{kong2016precision}, the authors put forward a graphical model that optimizes the expected performance measure and selectively displays facets just for part of queries by their predicted performance. 

After the rise of pre-trained language models, query facet generation has become a popular way of facet generation and achieved compelling performance. Hashemi et al. \cite{hashemi2021learning} proposed NMIR to cluster the documents prior to generation and learned a representation for each facet. Subsequently, to address the issue of matching between clustered documents and facets, as well as the influence caused by the order of generating facets, they proposed PINIMR \cite{hashemi2022stochastic}. \citet{samarinas2022revisiting} revisited the task of query facet extraction and generation and considered facet generation as autoregressive text generation which produces state-of-the-art results.

Inspired by the aforementioned works, in this paper, we summarize the impact of various training objectives on facet generation and conduct a comparative analysis of the generated results. We hope to provide guidance for the research in facet generation.

\section{METHODOLOGY}
In this section, we first introduce the definition of the facet generation task. We illustrate two existing training objectives in Section \ref{sec:facet generation without generated facets} and the newly proposed three in Section \ref{subsec:new_obj}.

\subsection{Task Description}
Given an open-domain query, our task is to generate the associated facets based on their corresponding related search engine result pages (SERPs). We use top-retrieved documents in the SERPs as evidence to help generate better facets during both training and inference. Let $D = \{(q_1,D_1,F_1), (q_2,D_2,F_2), \cdots, (q_n,D_n,F_n)\}$ denote the training data which consist of $n$ triples $(q_i,D_i,F_i)$, where $q_i$ means the $i-th$ open domain query, $D_i = \{d_{i1}, d_{i2}, \cdots, d_{ik}\}$ is the top $k_i$ retrieved documents for the given query $q_i$ and $F_i = \{f_{i1}, f_{i2}, \cdots, f_{im}\}$ represents $m_i$ ground truth facets related to $q_i$. The task is to generate a set of related facets $F$ for any given query $q$ with its associated documents $D$.

In this section, we describe five representative training objectives for facet generation. Two of them have been proposed in previous work and both of them conduct sequential facet prediction. They are: (1) \textit{seq-default} \cite{samarinas2022revisiting, hashemi2021learning} and (2) \textit{seq-min-perm} \cite{hashemi2022stochastic}. These two are order-sensitive and permutation-invariant respectively. Moreover, we propose another three permutation-invariant training objectives: (3) \textit{seq-avg-perm}, (4) \textit{set-pred}, and (5) \textit{seq-set-pred}. The comparative characteristics of these five objectives can be seen in Table \ref{tab:different objectives}. As in \cite{samarinas2022revisiting, hashemi2022stochastic, hashemi2021learning}, we use an autoregressive model BART \cite{lewis2019bart}, a Transformer-based encoder-decoder model defined by the parameter $\theta$ for sequence generation and leave the studies based on decoder-only methods in the future. Next, we will describe the details of all the objectives.

\subsection{Existing Objectives for Facet Generation \label{sec:facet generation without generated facets}}
In this subsection, we introduce two existing methods used for facet generation \textit{seq-default} and \textit{set-min-perm}.

\textbf{Seq-Default} \cite{hashemi2021learning, samarinas2022revisiting}. It considers the default facet sequences in the corpus as training targets and is commonly used in previous work \cite{hashemi2021learning, samarinas2022revisiting}. For a given query $q_i$ and its corresponding related documents $D_i$,
we concatenate $q_i$ and each $d_{ij} \in D_i$ using [SEP] and produce  $q_i[SEP]d_{i1}[SEP]\cdots[SEP]d_{ik}$ as input $x_i$. Note that we follow the same process of encoding this concatenated sequence with BART for all the studied training methods. We concatenate $f_{ij} \in F_i$ with `,' and the yielded text string  $f_{i1},f_{i2},\cdots,f_{im} [EOS]$ is the target $y_i$ for input $x_i$. This objective can be expressed in the following mathematical forms:
\begin{equation}
\theta=\arg \min _{\theta^*} \sum_{i=1}^n P(v,y_i),
\end{equation}
where $P(v,y_i) = \frac{1}{\left|y_i \right|}\sum_{x=1}^{\left|y_i \right|}-\log p\left(y_{i x} \mid v, y_{i 1}, \cdots, y_{i x-1} \right)$ and $v$ is the output of BART encoder. This training objective could lead to suboptimal results since the model learns towards only the given facet ordering and ignores other equally valid permutations. This could harm the model performance. Aware of this issue, Hashemi et al. \cite{hashemi2022stochastic} have proposed a loss function that is permutation-invariant, which we name as \textit{seq-min-perm} and will introduce next.

\textbf{Seq-Min-Perm} \cite{hashemi2022stochastic}. It treats the query intents as a set rather than a sequence to eliminate the impact of facet order. It extends the Hungarian loss \cite{kuhn1955hungarian} for facet generation. For a given input $x_i$, this method takes all the permutations of facets in $F_i$ into consideration. Each permutation is concatenated as a sequence and the sequence with the minimum loss is used as the training target. The way of concatenation to yield input is the same as \textit{seq-default} for a query $q_i$. The objective is as follows:

\begin{equation}
\theta=\arg \min _{\theta^*} \sum_{i=1}^n \min_{y_i^* \in \pi(y_i)} P(v, y_i^*),
\end{equation}
where the definition of $P(v, y_i^*)$ is the same negative log-likelihood function as \textit{seq-default}, $v$ is the output of BART encoder, $\pi(F_i)$ means all the possible permutations of ground truth facets for query $q_i$, and the number of items in the permutation set $\left|\pi(F_i)\right|$ equals to $m!$. \textit{Seq-min-perm} also has some limitations. Despite more computation on the permutations, only one permutation will have an impact on the model update, which could be insufficient. Also, the facet ordering with the minimum loss may be random in the early stages of model training. This random selection introduces noise to the training process and could result in a decrease in performance. \looseness=-1

For both \textit{seq-default} and \textit{seq-min-perm}, during inference, we greedily generate each word with the highest probability, expecting the model to automatically generate the separator `,' and termination symbol `</s>' to distinguish and end the generated facets. \textit{Seq-default} is order sensitive and \textit{seq-min-perm} is permutation-invariant. They both conduct sequential facet prediction and adaptively terminate the facet generation process. Also, \textit{seq-min-perm} has larger time complexity than \textit{seq-default}. 

\subsection{New Objectives for Facet Generation}
\label{subsec:new_obj}
To better model the facet generation process, we propose another three training objectives. They are all permutation-invariant but have different characteristics regarding whether to sequentially predict facets based on previously generated ones and whether the count of generated facets is controllable. We will describe each of them in detail next. 

\textbf{Seq-Avg-Perm}. It is a straightforward extension of \textit{seq-min-perm} that sequentially generates facets and is trained with the average loss of the permutations of facets. We permute the concatenation of facets in order to let the model learn towards all the possible permutations of the ground-truth facets which enhances its ability to search for the optimal solution. In particular, we employ the following training objective:
\begin{equation}
\theta=\arg \min _{\theta^*} \sum_{i=1}^n \frac{1}{\left|\pi(y_i)\right|}\sum_{y_i^* \in \pi(y_i)} P(v, y_i^*),
\end{equation}
where $P(v, y_i^*)$, $v$ and $\pi(y_i)$ are the same as \textit{seq-min-perm}. The inference procedure is also the same as \textit{seq-default} and \textit{seq-min-perm}. \looseness=-1

\textit{Seq-avg-perm} has the same time complexity as \textit{seq-min-perm} but instead uses all the samples for model updates. It also conducts sequential facet prediction so that previously generated facets could guide the generation of the next facet away from them. Same as \textit{seq-default} and \textit{seq-min-perm}, \textit{seq-avg-perm} also cannot generate more query facets when the final termination symbol is generated.  

\textbf{Set-Pred}. Instead of using the so-far predicted facets as contexts  for the current facet generation,  \textit{set-pred} treats each facet as an individual target and conducts parallel predictions. For a given query $q_i$ and its related documents $D_i$, we concatenate them and obtain the input $x_i$. In contrast to the other methods, we use an independent facet $f_{ij}$ as target output for the input $x_{i}$ and totally construct $m_i$ tuples $(x_i,f_{ij})$ for training, where $m_i$ is the number of facets that query $q_i$ has. So the optimization objective is:

\begin{equation}
\theta =\arg \min _{\theta^*} \sum_{i=1}^n  \frac{1}{m_i}\sum_{f_{ij} \in F_{i}} P(v, f_{ij}),
\end{equation}
where $v$ is the output of BART encoder and $P$ is again the negative likelihood probability of $f_{ij}$ given $v$. During inference, we generate multiple facets by taking the top z most likely predictions using beam search.

\begin{figure}[h]
  \centering
  \includegraphics[width=\linewidth]{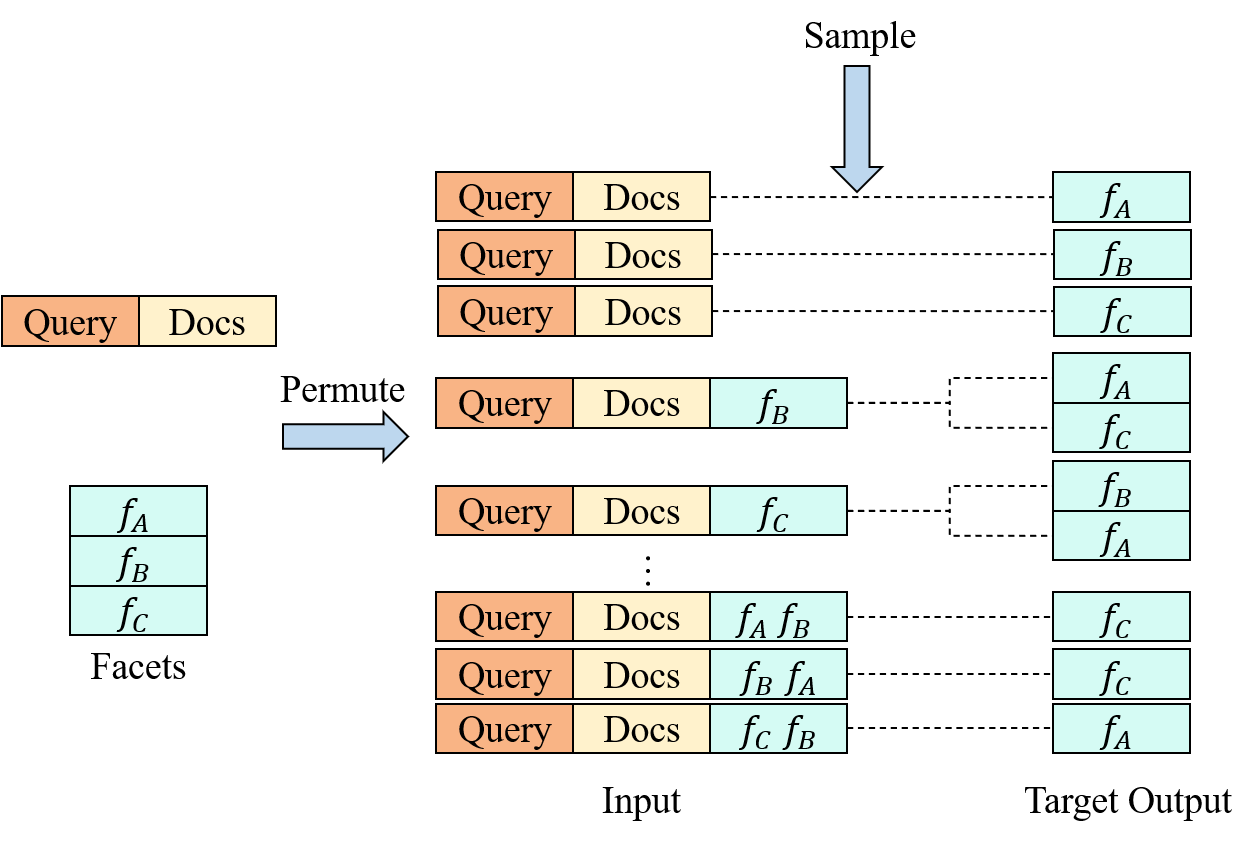}
  \caption{The organization of training data for seq-set-pred}
  \label{The organization of training data for FG-OD}
\end{figure}
The time complexity of \textit{set-pred} is much lower than \textit{seq-min-perm} and \textit{seq-avg-perm} while higher than \textit{seq-default}. According to \cite{vijayakumar2018diverse}, vanilla beam search could output synonyms in the top predictions. Similarly, the generated facets by \textit{set-pred} may be synonyms or refer to the same concept. To enhance the diversity of generated facets, we combine the ideas of \textit{set-avg-perm} and \textit{set-pred} and propose the \textit{seq-set-pred} objective. 

\textbf{Seq-Set-Pred}. It predicts each of the remaining facet sets in parallel based on arbitrarily generated facets as context. For a given query $q_i$, we divide the generation of facets into $\left|F_i\right|$ steps and generate one facet at each step. We concatenate the input $x_{t-1}$ with the output $y_{t-1}$ from the (t-1)-th step to form the t-th step's input $ x_t = x_{t-1} [SEP] y_{t-1}$.
During training, similar to \textit{seq-avg-perm}, to mitigate the influence of facet order, we train the model on the data that covers all the possible permutations. The organization of training data is depicted in Figure \ref{The organization of training data for FG-OD}. This prevents the model from learning specific generation ordering and enables it to make precise predictions under different permutation contexts effectively. We formalize this optimization  objective as:
\begin{equation}
\theta=\arg \min _{\theta^*} \sum_{i=1}^n \frac{1}{V_i}\sum_{j=0}^{\left|F_i \right|-1} \sum_{k=1}^{\left|\mathcal{A}_j(F_i)\right|} \sum_{h=1}^{\left|\mathcal{A}_1(R_{ijk})\right|}P(v_{ijk}, y_{ijkh}) ,
\end{equation}
where $P(v,y)$ is the negative log likelihood of the generation probability for $y$ given $v$. $\mathcal{A}_j(X)$ means all the possible orderings of $j$ selected non-repetitive elements from $X$. For example, $\mathcal{A}_2(\{f_A, f_B, f_C\}) = \{(f_A,f_B), (f_B,f_A), (f_A,f_C), (f_C,f_A), (f_B,f_C), (f_C,f_B)\}$. $R_{ijk}$ is the remaining set consisting of facets that are in $F_i$ but not in the k-th set in $\mathcal{A}_j(F_i)_k$. $v_{ijk}$ is the output of BART encoder given the input $x_{ijk}$, i.e., the concatenation of $q_i$, $D_i$ and facets in $\mathcal{A}_j(F_i)_k$. $y_{ijkh}$ is the h-th ground truth facet chosen from $R_{ijk}$. We get the average loss of all the possible permutations for query $q_i$ by dividing $V_i = {\sum_{j=0}^{\left|F_i \right|-1} \sum_{k=1}^{\left|\mathcal{A}_j(F_i)\right|}\left|\mathcal{A}_1(R_{ijk})\right|}$

The time complexity of \textit{seq-set-pred} is the highest among the five approaches. It also conducts sequential facet prediction so that the previously predicted facets could help it avoid generating similar facets. Same as \textit{set-pred}, it is able to control the count of generated facets. Note that based on decoder-only language generation models such as the GPT series \cite{radford2019gpt2,brown2020gpt3}, \textit{seq-set-pred} will be essentially the same as \textit{seq-avg-perm} since we do not need to move the generated facets to the encoder.  

\subsection{Model Inference}
For \textit{seq-avg-perm} inference, we generate facets the same as \textit{seq-default} and \textit{seq-min-perm}. For \textit{set-pred} and \textit{seq-set-pred}, we generate each facet independently and set the number of facets manually. We generate facets in parallel and utilize a search algorithm to select the top few facets with the highest probability as the generated results for \textit{seq-min-perm} while for \textit{seq-set-pred}, we sequentially generate all the facets, appending the generated facets to the current input as the input of next step, until the count of generated facets reaches the specified number.

\subsection{Stochastic Optimization}
Optimizing towards all permutations of query facets is computationally challenging. Therefore, as \cite{hashemi2022stochastic}, we adopt the stochastic variations of \textit{seq-min-perm}, \textit{seq-avg-perm}, and \textit{seq-set-pred}. In other words, instead of taking all permutations into consideration, we just sample a certain number of permutations from all the possible ones for each query and compute the approximate losses. We dynamically sample the permutations during each training epoch to make sure that the model can see as many different permutations as possible.

\begin{table*}[htbp]
  \centering
    \caption{Evaluation of different methods based on matching scores. The superscript $+$ denotes significant improvements compared to the worst one which is underscored and $-$ means significant decreases compared to the best one which is in bold in terms of a two-tailed paired t-test with Bonferroni correction with 99\% confidence.}
    \label{tab:matching score table}
  \begin{adjustbox}{width=\textwidth}
  \begin{tabular}{lccccccccccccc}
    \toprule
  \multirow{2}{*}{Model}
  & \multicolumn{3}{c}{Term Overlap} &  \multicolumn{3}{c}{Exact Match} & \multicolumn{4}{c}{Set BLEU Score} & \multicolumn{3}{c}{Set BERT-Score}\\
         \cmidrule(lr){2-4} \cmidrule(lr){5-7} \cmidrule(lr){8-11} \cmidrule(lr){12-14}
          & P & R & F1 & P & R & F1 & 1-gram & 2-gram & 3-gram & 4-gram & P & R & F1 \\
    \midrule
    Seq-Default & $0.2976^+$ & $0.2769^+$\rlap{$^-$} & $0.2752^+$\rlap{$^-$} & $0.0718^-$ & $0.0561^-$ & $0.0611^-$ & $0.2335^+$\rlap{$^-$} & $0.1040^+$\rlap{$^-$} & $0.0444^+$\rlap{$^-$} & $0.0175^+$\rlap{$^-$} & $\underline{0.6391}^-$ & $\underline{0.6455}^-$ & $\underline{0.6419}^-$\\
    Seq-Min-Perm & $\underline{0.2761}^-$ & $\underline{0.2536}^-$ & $\underline{0.2537}^-$ & $\underline{0.0620}^-$ & $\underline{0.0470}^-$ & $\underline{0.0519}^-$ & $\underline{0.2102}^-$ & $\underline{0.0850}^-$ & $\underline{0.0346}^-$ & $\underline{0.0125}^-$ & $0.6442^+$\rlap{$^-$} & $0.6482^-$ & $0.6457^+$\rlap{$^-$} \\
    \midrule
    Seq-Avg-Perm & $0.2977^+$ & $\textbf{0.3263}^+$ & $\textbf{0.3005}^+$ & $\textbf{0.1040}^+$ & $0.0960^+$ & $\textbf{0.0977}^+$ & $0.2422^+$\rlap{$^-$} & $0.1081^+$\rlap{$^-$} & $0.0568^+$\rlap{$^-$} & $0.0288^+$ & $0.6665^+$\rlap{$^-$} & $0.6697^+$\rlap{$^-$} & $0.6676^+$\rlap{$^-$}\\
    Set-Pred & $\textbf{0.3029}^+$ & $0.2978^+$\rlap{$^-$} & $0.2897^+$ & $0.0988^+$ & $0.0973^+$ & $0.0953^+$ & $0.2567^+$ & $0.1198^+$ & $0.0606^+$\rlap{$^-$} & $0.0260^+$\rlap{$^-$} & $\textbf{0.6873}^+$ & $\textbf{0.6897}^+$ & $\textbf{0.6880}^+$\\
    Seq-Set-Pred & $0.2930^+$ & $0.2989^+$\rlap{$^-$} & $0.2863^+$\rlap{$^-$} &$ 0.0993^+$ & $\textbf{0.1009}^+$ & $0.0973^+$ & $\textbf{0.2577}^+$ & $\textbf{0.1228}^+$ & $\textbf{0.0676}^+$ & $\textbf{0.0308}^+$ & $0.6849^+$ & $0.6887^+$ & $0.6863^+$ \\
    \bottomrule
  \end{tabular}
  \end{adjustbox}
\end{table*}
\section{EXPERIMENTAL Setup}
This section introduces the data we use for training and evaluation, the metrics we use to evaluate the models, and the technical details of the experiments.
\subsection{Dataset}
Following \cite{samarinas2022revisiting, hashemi2021learning}, our experiments are based on the MIMICS dataset \cite{zamani2020mimics}. MIMICS is a collection of search clarification datasets for real search queries sampled from the Bing query logs and it contains three subsets: MIMICS-Click, MIMICS-ClickExplore and MIMICS-Manual. For each search query, it provides up to 5 ground-truth facets and at most 10 associated documents with information such as document snippets. We use MIMICS-Click which includes over 400K unique queries for training and MIMICS-Manual which contains 2832 queries for evaluation. For all the training objectives, we use the document snippets provided by MIMICS as our document text.

\subsection{Evaluation Metrics \label{sec:evaluation metrics}}
We evaluate our approach in terms of two aspects: the matching between the generated facets and the ground-truth facets and the diversity among the generated facets.
On the one hand, to evaluate our approach in terms of matching with ground truth, we follow \citet{samarinas2022revisiting} and adopt four sets of evaluation metrics. (1) Precision, recall, and F1 of term overlap metrics: These metrics are computed based on the matching between the set of generated facet terms and the set of ground-truth facet terms at the term level. (2) Exact match: These metrics also compute precision, recall, and F1 between the facets generated by the model and the ground-truth facets, but at the facet level. (3) Set BLEU score: It calculates the BLEU \cite{papineni2002bleu} scores between the best permutation of generated facets and the ground-truth facets (4) Set BERT-Score: It calculates the BERTScore \cite{zhang2019bertscore}  between the best permutation of generated facets and the ground-truth facets. For more details of the metrics, please refer to \cite{hashemi2021learning}.

On the other hand, we propose two extra metrics to measure the diversity of a set of facets. (1) Term diversity: For a given set $A = \{A_1, A_2, \cdots, A_n\}$ where $A_i$ means the i-th facet in $A$, we calculate the term-level diversity of $A$ with the average of one minus the overlap ratio between each pair of facets in $A$ and the overlap ratio between $A_i$ and $A_j$ is computed by $2 * \frac{\left|A_j\cap A_j\right|}{\left|A_i\right| + \left|A_j\right|}$. (2) BERT-score diversity: for the set $A$, we calculate the average BERTScore between every pair of facets in the set where BERTScore computes a similarity score for each token in the candidate sentence with each token in the reference sentence and the token similarity is computed using contextual embeddings. We use one minus it as the BERT-score diversity.

\subsection{Implementation Details}
We fine-tuned BART-base for five epochs with an initial learning rate set to $5 \times 10^{-5}$ for all the following approaches as \cite{samarinas2022revisiting} and employed beam search algorithm with a beam size of 5. If not specifically mentioned, we use AdamW optimizer and set the maximum sequence length to 512 tokens, the maximum output length to 32 tokens, and batch size to 16. The count of generated facets for \textit{seq-default, seq-min-perm,} and \textit{seq-avg-perm} is determined by themselves, while we specifically set the number of generated facets to 3 for \textit{set-pred} and \textit{seq-set-pred}, as it yields the best results on the validation set. Additionally, practical considerations led us to perform deduplication on all the generated results for each method. It is worth noting that there was only a minor difference between the results before and after deduplication. For \textit{seq-default}, we utilized the checkpoint provided by Samarinas et al. \cite{samarinas2022revisiting}.

For the \textit{seq-avg-perm} and \textit{seq-min-perm} techniques, we also fine-tuned the BART-base model with a maximum sequence length of 512 tokens. However, we augmented the maximum output sequence length to 128 tokens to empower the model with the capability to generate multiple facets within a single sequence. During each training epoch, we sampled six permutations for each query and the batch size for \textit{seq-min-perm} was set to 18, enabling it to handle all permutations of three queries within one batch. 

Regarding \textit{seq-set-pred}, we set the maximum input sequence length to 640 tokens. This ensured that the facets appended to the input would not be truncated. We sampled 6 permutations, 8 permutations, 9 permutations, 11 permutations, and 13 permutations when the count of ground-truth facets is 1, 2, 3, 4, 5 respectively to ensure the model to see almost the same number of facets as in the \textit{seq-avg-perm} approach during each training epoch. 

\section{Results and Discussion}
Next, we show the experimental results of the five training objectives. First, we evaluate the accuracy of the generated facets against the ground-truth facets. Then, we measure their diversity with our proposed diversity metrics. We also show the model performance when different counts of facets are generated and compare the model performance using a similar amount of training data. Finally, we conduct case analyses to show the quality of the facets generated by each method. 

\begin{table}[htbp]
\caption{Diversity evaluation on the facet body which removes words from the facets that appear in the query.}
\label{tab:diversity}
\setlength\tabcolsep{5.5pt}
  \begin{tabular}{lcccc}
    \toprule
     \multirow{2}{*}{Model} & Term Diversity &  \multicolumn{3}{c}{BERT-Score Diversity}\\
     & Ratio & P & R & F1 \\
    \midrule
    Ground Truth & 0.9284 & 0.0829 & 0.0829 & 0.0836 \\
    \midrule
    Seq-Default & 0.8783 & 0.0743 & 0.0743 & 0.0749 \\
    Seq-Min-Perm & 0.8922 & 0.0883 & 0.0889 & 0.0893  \\
    Seq-Avg-Perm & \textbf{0.9218} & \textbf{0.0993} & \textbf{0.0989} & \textbf{0.1000} \\
    Set-Pred & 0.8883 & 0.0630 & 0.0630 & 0.0635 \\
    Seq-Set-Pred & 0.9117 & 0.0657 & 0.0666 & 0.0667 \\
    \bottomrule
  \end{tabular}
\end{table}

\subsection{Evaluation Against Ground Truth}
Table \ref{tab:matching score table} shows the matching degree between the generated facets and ground-truth facets for all the methods.
We have the following observations: 1) Most permutation-invariant methods except \textit{seq-min-perm} perform better than the order-sensitive methods. This is consistent with our presumption that \textit{seq-default} was hurt when forced to generate a specific facet ordering that is only one of the possible combinations. However, \textit{seq-min-perm} is the exception and performs the worst most of the time. There are some potential reasons for its unsatisfactory performance. 
If the model consistently selects the same sequence for a query throughout the entire training phase, it is expected to observe a similar performance to that of \textit{seq-default}. However, in the early stages of training, the selection of the sequence with the minimum loss exhibits randomness. This random selection introduces noise to the training process, resulting in a decrease in performance. So the effectiveness of generated facets is even worse than \textit{seq-default}. 
2) The method that generates facets without depending on the previously generated facets (i.e., \textit{set-pred}) has compelling performance in terms of both term-based and semantic matching metrics. Previous studies do not consider this way of training so this has not been observed. Its diversity, however, is lower than the others, which we will show in Section \ref{subsec:diversity}. 
3) Methods that only learn facet prediction given context(e.g., \textit{seq-set-pred}) have better semantic matching performance with ground truth compared to those that also learn when to stop generating facets. \looseness=-1



\subsection{Evaluation on Diversity}
\label{subsec:diversity}
Table \ref{tab:diversity} shows the results of diversity evaluation on the generated facets with the query words removed. It indicates that \textit{seq-avg-perm} performs the best on both metrics, suggesting there are not only fewer repeated terms between the generated facets but also a higher level of semantic differentiation. It performs worse than the ground-truth facets regarding term diversity but better in terms of semantic diversity. By checking some of their generated facets, we find that \textit{seq-avg-perm} usually generates terms such as prepositions to maintain correct grammatical structures while such connection words are fewer in the ground truth and this phenomenon also exists in other methods. For example, given query "internet explorer", \textit{seq-avg-perm} generates facets "for windows 10" and "for windows 7" while the corresponding ground truth facets are actually "windows 10" and "windows 7" respectively(shown in Table \ref{tab:case}). This leads to a decrease regarding term diversity for \textit{seq-avg-perm}. The higher semantic diversity scores of \textit{seq-avg-perm} compared to the ground truth indicates that its generated facets are more semantically different.
It is worth noting that all the sequential prediction methods, except for \textit{seq-default}, exhibit higher diversity than \textit{set-pred}. It is not surprising to observe worse diversity in \textit{set-pred} because it could keep generating similar facets since it only refers to the original query and top documents during generation. The possible reason why \textit{seq-default} performs poorly regarding term diversity is that it is trained towards only one possible permutation of the facets, which could constrain the output term space and in turn result in lower term diversity. We observe good term diversity but worse semantic diversity in \textit{seq-set-pred}. Multiple target ground-truth facets could be helpful for it to obtain higher term diversity. However, when we move the previously generated facets to the encoder and predict the rest with the decoder, the encoder is struggling to learn that the target facet should be similar to the original query while different from the facets. In contrast, for the sequential prediction methods(i.e., \textit{seq-default, seq-min-perm, seq-avg-perm}), the encoder only encodes query and captures its relevance with the target facet and it is feasible for the decoder to capture the difference between target facet with the previously generated facets. 
This could be the reason for the much lower semantic diversity of \textit{seq-set-pred} compared to the sequential prediction counterparts. 

\begin{table}[htbp]
\caption{Matching scores and diversity on a different number of facets. Div means diversity for short.}
\label{tab:control facet number results}
\scalebox{0.73}{
  \begin{tabular}{lccc|cc}
    \toprule
   \multirow{2}{*}{Model} & & Term Overlap  & Set BERT-Score & Term Div & BERT-Score Div\\
    & Num & F1 & F1 & Ratio & F1\\
    \midrule
    \multirow{5}{*}{Set-Pred} & 1 & 0.2696 & 0.3254 & - & -  \\
     &2 & 0.2888 & 0.6477 &  0.8812 & 0.0613  \\
     &3 & \textbf{0.2897} & \textbf{0.6880} & 0.8883 & 0.0635   \\
     &4 & 0.2867 & 0.6231 & \textbf{0.8903} &  0.0658\\
     &5 & 0.2814 & 0.5450 & 0.8869 & \textbf{0.0683}  \\
     \midrule
     \multirow{5}{*}{Seq-Set-Pred} & 1 & 0.2709 & 0.3253 & - & -  \\
     &2 & \textbf{0.2886} & 0.6474 & \textbf{0.9133} & 0.0649  \\
     &3 & 0.2863 & \textbf{0.6863} & 0.9117 & 0.0667 \\
     &4 & 0.2759 & 0.6259 & 0.9095 & 0.0687 \\
     &5 & 0.2677 & 0.5660 & 0.9083 & \textbf{0.0705}\\
    \bottomrule
  \end{tabular}
  }
\end{table}

\begin{table}[htbp]
\caption{The proportion of generated facets that match the ground truth facets.}
\label{tab:ratio}
\scalebox{0.78}{
  \begin{tabular}{lccccc}
    \toprule
      & Seq-default & Seq-Min-Seq & Seq-Avg-Perm & Three Facets & Two Facets\\
    \midrule
    Ratio & 0.7038 & 0.7072 & 0.6678 & 0.7039 & \textbf{0.7431}\\
    \bottomrule
  \end{tabular}
  }
\end{table}


\subsection{Performance w.r.t. Facet Counts}
Table \ref{tab:control facet number results} shows the performance when generating different numbers of facets using the facet-count-controllable methods \textit{set-pred} and \textit{seq-set-pred}. We also compute the ratio of the generated facet counts matching the ground truth for each method and show the ratios in \ref{tab:ratio}.  Table \ref{tab:control facet number results} shows that the evaluation scores constantly change with different facet counts. 
The best matching scores are mainly achieved when generating 2 or 3 facets because the average number of ground truth facets is between 2 and 3(shown in Table \ref{tab:ratio}). The ratio is the average of one minus $\frac{\left|f_i-g_i\right|}{g_i}$ across all the queries where $g_i$ is the count of ground-truth facets for $q_i$. The ratio matching with the ground-truth facet counts is 0.7039 and 0.7431 when generating 3 and 2 facets respectively. Although \textit{seq-default} and \textit{seq-min-perm} have generated more facets that have the same count with ground truth than \textit{seq-avg-perm}, they have worse term or semantic level matching scores with the ground-truth facets. It indicates that they do generate worse facet contents. \textit{Set-pred} and \textit{seq-set-pred} have similar accuracy in terms of facet counts than \textit{seq-default} and \textit{seq-min-perm} but have better facet contents as well. Compared to \textit{seq-avg-perm}, facet-count-controllable methods perform better on set BLEU score and set BERT-Score. The possible reason is that these two metrics are more sensitive to the number of generated facets. When calculating these two metrics, facets that do not match the ground-truth facets in the count will receive a score of 0 and the mismatch between the count of generated facets and the count of ground-truth facets results in worse matching scores. However, this phenomenon does not exist When calculating BERT-score diversity. Because each facet has its comparable facets.

When it comes to diversity, the term diversity and BERT-score diversity of \textit{seq-set-pred} decrease and increase, respectively, with the increase in the number of generated facets. The results of \textit{set-pred} indicate that this approach performs well when generating a smaller number of facets. However, when generating more facets, it may produce more repeated tokens. In conclusion, \textit{seq-set-pred} demonstrates better diversity than \textit{set-pred} across all the numbers of generated facets.

\begin{figure}[htbp]
  \centering
  \includegraphics[width=\linewidth]{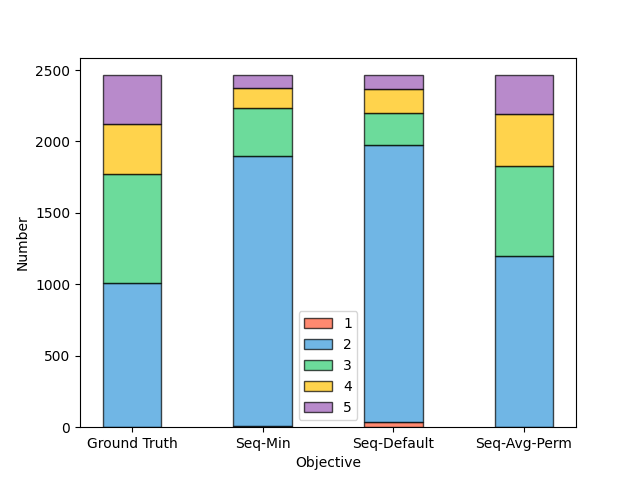}
  \caption{Distribution of the number of facets generated by different models.}
  \label{fig:number}
\end{figure}

Simultaneously, we visualize the count of facets generated by the adaptive generation methods as Figure \ref{fig:number}. It demonstrates that \textit{seq-default} and \textit{seq-min-perm} tend to generate more results with two facets while \textit{seq-avg-perm} generate various numbers of facets.

\begin{table*}[htbp]
\caption{Some examples of the facets generated by each model. We selected the top-5 facets for Set-Pred and Seq-Set-Pred. The duplicate facets were removed from all the models. Facets are separated using the `,' symbol.}
  \label{tab:case}
  \scalebox{0.95}{
  \begin{tabular}{lcccc}
    \toprule
    \multirow{2}{*}{Model} & \multicolumn{3}{c}{Query}\\
    & internet explorer & nike boys shoes & abiotic factors \\
    \midrule
    \multirow{2}{*}{Ground Truth} &  windows 10, windows 7, & basketball shoes, running shoes, & grasslands, savanna, \\
    & windows 8, windows xp & tennis shoes, soccer shoes, golf shoes & tundra \\

    \midrule
    \multirow{2}{*}{Seq-default} & internet explorer windows 7, & nike boys tennis shoes, & tropical rainforest, temperate forest, \\
    & internet explorer windows 10 & nike boys running shoes & tundra, desert, savanna \\

    \midrule
    \multirow{2}{*}{Seq-Min-Perm} & internet explorer 32 bit, & nike boys shoes for boys, & in animals, \\
    & internet explorer 64 bit & for men, for kids & for kids \\
    
    \midrule
    \multirow{2}{*}{Seq-Avg-Perm} & internet explorer for windows 10, windows xp, & running shoes, basketball shoes, & taiga, savanna. \\
    &internet explorer windows 8 , for windows 7 & golf shoes, tennis shoes, soccer shoes & tundra, desert \\

    \midrule
    \multirow{5}{*}{Set-Pred} & for windows vista, & nike boys basketball shoes, & of tundra,\\
    & for windows 10, & nike boys golf shoes, & of the rainforest, \\ 
    & for windows 8,  & nike boys tennis shoes, & tropical rainforest, \\
    & for windows 7 ,& nike boys running shoes, & tundra, \\
    & for windows xp & basketball shoes & tundra habitat \\

    \midrule
    \multirow{5}{*}{Seq-Set-Pred} & internet explorer windows 7, & nike boys basketball shoes,  & in desert,\\
    & internet explorer windows xp, & nike boys running shoes, & in rainforest, \\ 
    & internet explorer windows 10,  & nike boys baseball shoes, & in the tundra, \\
    & internet explorer windows server 2012, & nike boys football shoes, & in the savanna \\
    & internet explorer windows 8 & & \\
    \bottomrule
  \end{tabular}
  }
\end{table*}

\begin{table}[htbp]
\caption{Matching scores of different methods based on similar training data amount. The superscript $+$ denotes significant improvements compared to the worst one which is underscored and $-$ means significant decreases compared to the best one which is bold in terms of a two-tailed paired t-test with Bonferroni correction with 99\% confidence.}
\label{tab:similar_quantity}
  \begin{tabular}{lcccc}
    \toprule
    & Term Overlap &  Exact Match & Set BERT-Score\\
    Model  & F1 & F1 & F1\\
    \midrule
    Seq-Default  & $0.2752^+$\rlap{$^-$} & $0.0611^+$\rlap{$^-$} & $\underline{0.6419}^-$ \\
    Seq-Min-Perm & $\underline{0.2537}^-$ & $\underline{0.0519}^-$ & $0.6457^+$\rlap{$^-$} \\
    \midrule
    
    Seq-Avg-Perm & $\textbf{0.2898}^+$ & $0.0737^+$\rlap{$^-$} & $0.6562^+$\rlap{$^-$} \\
    Set-Pred  & $0.2897^+$ & $\textbf{0.0953}^+$ & $\textbf{0.6880}^+$ \\
    Seq-Set-Pred & $0.2777^+$\rlap{$^-$} & $0.0816^+$\rlap{$^-$} & $0.6791^+$\rlap{$^-$} \\
    \bottomrule
  \end{tabular}
\end{table}

\begin{table}[htbp]
\caption{Comparison between ChatGPT with our studied methods on 50 random test samples in MIMICS-Manual. The worst performance is underscored and the best one is in bold.}
\label{tab:compare_chatgpt}
  \begin{tabular}{lccc}
    \toprule
    & Term Overlap &  Exact Match & Set BERT-Score\\
    Model  & F1 & F1 & F1\\
    \midrule
    Seq-Default  & $0.2566$ & $0.0374$ & $0.6070$ \\
    Seq-Min-Perm & $0.2627$ & $\underline{0.0044}$ & $0.6293$ \\
    Seq-Avg-Perm & $\textbf{0.3276}$ & $0.0760$ & $0.6654$ \\
    Set-Pred  & $0.3187$ & $\textbf{0.1225}$ & $\textbf{0.7039}$ \\
    Seq-Set-Pred & $0.2998$ & $0.1003$ & $0.5811$ \\
    \midrule
    ChatGPT & $\underline{0.1598}$ & 0.0315 & $\underline{0.5711}$ \\
    \bottomrule
  \end{tabular}
\end{table}

\subsection{Impact of Training Data Amount}
Due to the significant differences in the quantity of information utilized per training epoch for each method, we compare each method when the amount of training data is at a similar scale. Specifically, we only train the methods that learn towards all the facet permutations (\textit{seq-avg-perm} and \textit{seq-set-pred}) for 1 epoch, so that the overall training data amount is similar to the other methods. As shown in Table \ref{tab:similar_quantity}, their results have different extents of regressions since it takes longer for the model training to converge under the permutation-invariant context. However, both \textit{seq-avg-perm} and \textit{seq-set-pred} still outperform \textit{seq-default} and \textit{seq-min-perm} in terms of all the metrics. 

\subsection{Facet Generation with ChatGPT}
\begin{figure}[h]
  \centering
  \includegraphics[width=\linewidth]{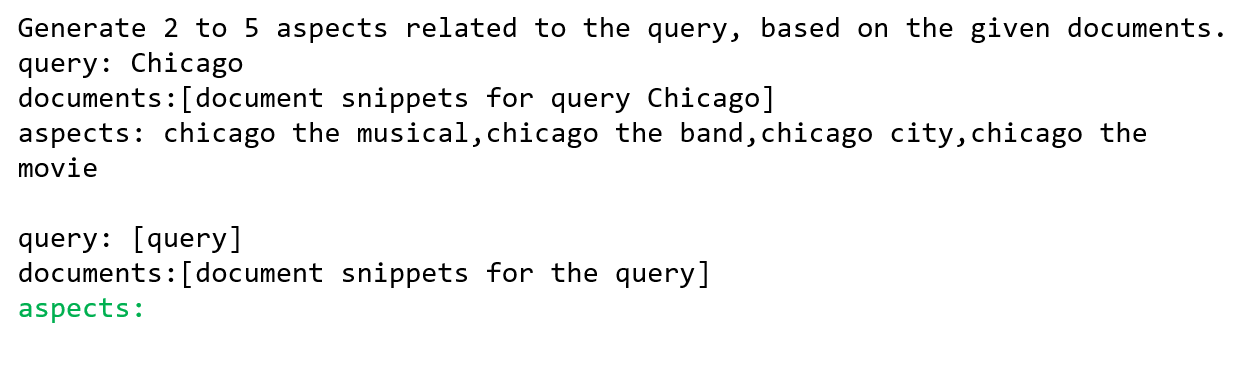}
  \caption{The prompt used for ChatGPT}
  \label{fig:prompt}
\end{figure}

Recently, Large Language Models (LLMs), such as ChatGPT, have demonstrated remarkable capabilities across various tasks. So, one may wonder how such LLMs perform on the facet generation task. With this regard, we assess the ability of ChatGPT in facet generation and compare it to the methods investigated in this paper. We randomly sampled 50 test examples from MIMICS-Manual dataset and ash ChatGPT to generate facets using the instruction as in Figure \ref{fig:prompt}. The results are displayed in Table \ref{tab:compare_chatgpt}. 
It shows that ChatGPT has a large performance gap compared to most of the methods presented in our paper across all metrics. However, despite ChatGPT's poor performance on the metrics, we still believe that its generated results are reasonable. For example, for the query ``Express vp'', the facets generated by ChatGPT are ``virtual private network'', ``privacy and security'', ``content access'', ``server network'', and ``pricing'' while the ground-truth facets are ``expressvpn mac'', ``expressvpn android'', ``expressvpn windows'', ``expressvpn linux'', and ``expressvpn ios''. This finding is consistent with the results in \cite{samarinas2022revisiting}. In sum, the facets generated by ChatGPT are general concepts related to the query but do not match the facets manually labeled according to the provided document snippets.

\subsection{Case Study}
We demonstrate the generated results for three queries with varying numbers of ground-truth facets in Table \ref{tab:case}. We can observe that \textit{seq-default} and \textit{seq-min-perm} tend to generate two facets for a given query, which aligns with the pattern shown in Figure \ref{fig:number}. Many facets generated by \textit{seq-default} match the terms in the ground truth, resulting in higher precision in term overlap, but due to limitations in the number of generation facets, the recall metric is not very high. For \textit{seq-min-perm}, the generated facets are somewhat related to the query but could not generate the desired facets, leading to lower scores. These findings are consistent with the results in Table \ref{tab:matching score table}. The remaining three methods \textit{set-pred, seq-avg-perm}, and \textit{seq-set-pred} demonstrate good generation performance. In the facet-count-controllable methods, we select the top three facets as the generated results. However, we can find that the ignored facets may be better. Therefore, we can generate more facets and choose the best facets instead of directly selecting the first three based on the similarity between the facets and the query. We will investigate this in the future. 

\section{CONCLUSIONS AND FUTURE WORK}
In this paper, we conducted a systematic comparative study of various types of training objectives, with different properties of, whether it is permutation-invariant, whether it conducts sequential prediction, and whether it can control the count of output facets.   For comprehensive comparisons, besides the commonly used evaluation that measures the matching with ground-truth facets, we also introduce two diversity metrics. Our experimental analyses demonstrate: the appropriate permutation-invariant objectives can help generate better facets; facet prediction that is only based on the query and
top-retrieved documents (i.e., \textit{Set-Pred}) achieves compelling performance in terms of the metrics measuring matching with the ground truth but have the worst diversity performance; methods that only learn facet prediction given context (i.e., \textit{Seq-Set-Pred}) have better semantic matching metrics with ground truth but worse diversity performance than its counterpart that also learns when to stop generating facets. Our newly proposed methods outperform the previous state-of-the-art models \cite{samarinas2022revisiting}.

For future work, We plan to evaluate these objectives on a decoder-only architecture such as GPT-2 \cite{radford2019gpt2} in the next step. As we mentioned in Section \ref{subsec:new_obj}, \textit{seq-perm-avg} and \textit{seq-set-pred} will be essentially the same based on the decoder-only models. Another interesting research direction is to utilize a small amount of data to learn how to predict user intents. Although MIMICS have enough annotations for fine-tuning, in reality it is common to have limited annotated data. Thus, we plan to study intent prediction with few-shot learning.

\begin{acks}
This work was funded by the National Natural Science Foundation of China (NSFC) under Grants No. 62302486, the Lenovo-CAS Joint Lab Youth Scientist Project, and the project under Grants No. JCKY2022130C039. Any opinions, findings, and conclusions or recommendations expressed in this material are those of the authors and do not necessarily reflect those of the sponsor.
\end{acks}

\bibliographystyle{ACM-Reference-Format}
\balance
\bibliography{reference}

\end{document}